\documentclass[12pt,epj]{revtex4}
\usepackage{fancyhdr}
\usepackage{amssymb}
\usepackage{amsmath}
\usepackage[mathscr]{eucal}
\usepackage{latexsym}
\usepackage{amsbsy}
\usepackage{float}
\usepackage[dvips]{graphicx}
\usepackage{epsfig}
\usepackage{subfigure}
\usepackage{wrapfig}
\usepackage{epsf}
\usepackage{float}

\newcommand{\be}{\begin{equation}}
\newcommand{\ee}{\end{equation}}
\newcommand{\bea}{\begin{eqnarray}}
\newcommand{\eea}{\end{eqnarray}}
\newcommand{\ba}{\begin{array}}
\newcommand{\ea}{\end{array}}

\newcommand{\Tr}{{\mathrm{Tr}}}

\begin{document}
\title{Anomalous dimensions of operators without derivatives in the non-linear $\sigma$-model for disordered bipartite lattices}

\author{Luca Dell'Anna}
\address{Max-Planck-Institut f\"ur Festk\"orperforschung, D-70569
Stuttgart, Germany}
\address{Institut f\"ur Theoretische Physik, Heinrich-Heine-Universit\"at, D-40225 D\"usseldorf, Germany.}
\maketitle
%

We consider a generic time-reversal invariant model of fermions hopping randomly 
on a square lattice. 
By means of the conventional replica-trick within the fermionic path-integral formalism,  
the model is mapped onto a non-linear $\sigma$-model with fields spanning the coset  
$U(4N)/Sp(2N)$, $N\to 0$.  
We determine the proper scaling combinations of an infinite family
of relevant operators which control deviations from perfect two-sublattice 
symmetry. This allows us to extract the low-energy behavior of the 
density of states, which agrees with earlier results obtained in particular 
two-sublattice models with Dirac-like single-particle dispersion.  
The agreement proves the efficacy of the conventional fermionic-path-integral 
approach to disordered systems, which, in spite of many controversial aspects, 
like the zero-replica 
limit, remains one of the more versatile theoretical tool to deal with disordered electrons. 
%
\maketitle

\section{Introduction}

It is known that localization does not occur in any dimension at the
band-center energy of tight binding models on bipartite lattice-Hamiltonians 
\cite{Theo,Egg,Weg,G&W,Gade,balents,Gogolin,Fabrizio,guruswamy,brouwer,ryu}. 
Gade and Wegner \cite{G&W,Gade} first realized that these models correspond to a
particular class of non-linear $\sigma$-models in the zero replica limit, 
so called two-sublattice models, which exhibit an additional chiral 
symmetry\cite{G&W,Gade,Fabrizio}. As a consequence of this symmetry,  
they were able to prove that, when the chemical potential is right in the centre of the band, 
quantum-interference corrections to the $\beta$-function 
vanish exactly. In addition they showed that, unlike conventional 
disordered systems, the density of states (DOS) $\rho(E)$ near the band-center $E=0$ 
is strongly affected by disorder. They actually predicted a diverging behavior 
\cite{G&W,Gade} (see also Refs.\cite{Fabrizio,guruswamy})
\[
\rho(E) \sim \frac{1}{|E|}\, {\rm e}^{\displaystyle -A\sqrt{\ln|B/E|}}
\]
where $A$ and $B$ are positive constants.  

The subleading dependence $\exp\big(-A\sqrt{\ln|B/E|}\big)$ 
has been recently questioned by Motrunich, Damle and Huse \cite{Damle}. They analysed 
the strong disorder limit of a set of models which belongs to the two-sublattice class 
and found that the correct subleading dependence is instead 
$\exp\big(-A|\ln B/E |^{2/3}\big)$. This result was later confirmed by 
field-theoretical approaches based on supersymmetry \cite{Mudry} 
and on replica trick \cite{yamada} applied to the so-called 
Hatsugai-Wen-Kohmoto (HWK) model, which describes electrons hopping randomly on 
a square lattice in the presence of a $\pi$-flux per plaquette. This model is 
particularly suitable for a weak-disorder field-theoretical approach. Indeed, 
for uniform hopping, the low-energy single-particle spectrum of the HWK model is composed by 
two Dirac-like cones, which allows the use of the full machinery of Conformal Field Theory 
when a weak random-hopping component is included. The important breakthrough put forward by 
these analyses is that many disorder-average quantities, like the density of states, 
are determined by an infinite set of relevant local operators with negative 
dimensions \cite{Mudry}. 

In reality, these new results raise an intriguing question about the 
concept of universality commonly accepted in disordered systems, according to which 
the HWK model should be representative of any two-sublattice model since the action of its 
long-wavelength diffusive modes is a non-linear $\sigma$-model 
in the same universality class as generic two-sublattice models. 
However the results found by Mudry, Ryu, and 
Furusaki \cite{Mudry} and by Yamada and Fukui \cite{yamada} have been obtained 
working directly with the HWK Hamiltonian without going through the non-linear $\sigma$-model 
mapping, just thanks to the Dirac-like dispersion. Lacking an independent derivation 
starting from the non-linear $\sigma$-model, it is not a priori obvious to what extent 
these results are actually generic to any two-sublattice model.   

In this work, we are going to show that  
the results of Refs.~\cite{Damle,Mudry,yamada} can be fully recovered through the 
conventional non-linear $\sigma$-model approach based on the replica-trick within the 
fermionic path-integral formalism, without assuming any Dirac-like dispersion. 
Besides satisfying a purely theoretical curiosity, this result proves once again the strength  
of the conventional approach to disordered systems.

The paper is organized as follows. In Section II we introduce the model as well as 
the non-linear $\sigma$-model which describes its 
long-wavelength diffuse modes. 
In Section III we analyse the scaling behavior 
of some operators which are compositions of such diffuse modes, 
called soft operators, and whose $\beta$-functions are solved in Section IV. 
Finally in Section V we apply what found in the previous sections to calculate 
the mean density of states near the center of the band.

\section{The model}

A two-sublattice model is described by  
a non-interacting Hamiltonian on a bipartite lattice, with sublattices $A$ and $B$,
of the general form 
\be
\mathcal{H} = -\sum_{R\in A}\,\sum_{R'\in B} t_{RR'}\,c^\dagger_{R}
c^{\phantom{\dagger}}_{R'} + H.c.,
\label{h}
\ee   
with random matrix elements $t_{RR'}$ which only connect one sublattice to the other.  
It follows that, if $\Psi(R)$ is an eigenstate with eigenvalue $E$, then 
the wavefunction $\phi(R)\,\Psi(R)$, where $\phi(R)=1$ if $R\in A$ 
and $\phi(R)=-1$ if $R\in B$, is also an eigenstate with eigenvalue $-E$. This also implies 
that any eigenstate at $E=0$ is doubly degenerate, unless boundary conditions break the 
degeneracy between the two sublattices.  
Following Ref.~\cite{Efetov}, we introduce, within the fermionic path-integral formalism, 
the following Nambu spinors  
\[
\psi_R=\frac{1}{\sqrt{2}}\left(\ba{l} 
\bar{c}_R\\c_R 
\ea\right)
\]
where $c_R$ and $\bar{c}_R$ are Grassmann variables with components $c_{R,p,a}$ and
$\bar{c}_{R,p,a}$ in which $R$ 
refers to a lattice site, $p=\pm$ is the index of positive or negative
frequency components, $a=1,..,N$ is the replica index which has to be sent to 
zero at the end of the calculation. In addition we define conjugate fields through  
\[
\bar{\psi}_R = \left[\hat{c}\,\psi_R\right]^t = \frac{1}{\sqrt{2}}
\big(-c_R,\bar{c}_R\big),
\]
where $\hat{c}$ is the charge conjugation matrix $\hat{c}=-i\tau_2$. Here and what follows
the $\tau_i$'s, $i=1,\dots,3$, are Pauli matrices acting in the Nambu space.
With these fields the action describing the model  
at a fixed disorder configuration for fields with energy $E\pm i\omega$ is 
\be
\label{act}
S=-\sum_{R\in A}\,\sum_{R'\in B}\, 
\bar{\psi}_R\big(E\delta_{RR'}-i\,\omega s_3\delta_{RR'}-t_{RR'}\big)\psi_R,
\ee
where $s_3$ is the third Pauli matrix acting on the two 
frequency components $E\pm i\omega$ and the frequency transferred $2\omega$ plays the role of 
a symmetry breaking field. Since the detailed derivation of the non-linear $\sigma$-model 
starting from the action (\ref{act}) is known, see for instance Ref.~\cite{Fabrizio}, here 
we just outline the main steps emphasizing the peculiarity of two-sublattice models. 
We start noticing that the action when $E=\omega=0$ is invariant under the transformation
\[
\psi_R \to {\rm e}^{i\alpha \phi(R)}\,\psi_R,\qquad 
\bar{\psi}_R \to \bar{\psi}_R\, {\rm e}^{i\alpha \phi(R)},
\]
which is allowed within the path-integral formalism since the Grassmann variables $c_R$ and 
$\bar{c}_R$ are independent. This additional abelian symmetry plays an important 
role in these models, as originally recognized by Gade and Wegner \cite{G&W}. 
Within the replica-trick technique, the average of disorder can be performed directly on the 
action (\ref{act}) and generates a non-local interaction which connects two sites belonging 
to different sublattices. This interaction is then decoupled 
by an Hubbard-Stratonovich transformation, introducing an auxiliary field $Q$. Due to 
the non-locality of the interaction, in the long-wavelength limit this field has two 
components, one {\sl uniform}, $Q_0(R)$, and the other {\sl staggered}, $\phi(R) Q_3(R)$, 
where both $Q_0$ and $Q_3$ vary smoothly in space. In particular, if 
we introduce the Pauli matrices $\gamma_i$, $i=1,2,3$, as well as the identity matrix 
$\gamma_0$ in the two-sublattice space, the action of $\gamma_3$ being just that of $\phi(R)$, 
the auxiliary field to which the electron density is coupled is 
$Q(R)=Q_0(R)\,\gamma_0 + i Q_3(R)\, \gamma_3$.
Both $Q_0$ and $Q_{3}$ are $4N\times 4N$ hermitian matrices in Nambu, 
energy and replica spaces [$4N=2$(Nambu spinor dimension)$\times 2$(frequency components)
$\times N$(replicas)].
The derivation proceeds then through the following steps: i)
integrating over the Grassmann variables, ii) expanding the effective
action around the symmetry breaking saddle point $Q_0=\Sigma s_3$, where $\Sigma$ is the
inverse relaxation time,
iii) integrating out massive modes and just focusing on 
low-energy long-wavelength transverse fluctuations. In this way 
one obtains the following effective action for the auxiliary field when $E=0$
\cite{Gade,Fabrizio},
\be
\label{F0}
F_{_0}=\int dr \left[ \frac{\pi\sigma}{4^2}\,\Tr \left(\nabla Q^{\dagger}(r)\nabla Q(r)\right) 
-\frac{\pi\Pi}{2\cdot4^3}\,\left(\Tr\left(Q^{\dagger}(r)\nabla Q(r)\gamma_3\right)\right)^2 - \frac{\pi\nu \omega}{4}\,\Tr \left(s_3 Q(r)\right)\right],
\ee
In the above equation we have rescaled $Q\rightarrow \Sigma Q$ 
with $Q=\widetilde U^\dagger s_3 U$ 
and the unitary transformation $U\in U(4N)/Sp(2N)$ fulfilling
\be 
\label{tildeU}
\widetilde U^\dagger \equiv \gamma_1\, U^\dagger\, \gamma_1=\hat{c}\, U^t \hat{c}^t.
\ee 
The latter relation implies the constrain $Q^{\dagger}Q=1$. 
The coupling $\sigma$ is
the conductivity in Born approximation, $\Pi$ is related to the staggered
density of states fluctuations \cite{Fabrizio} and $\nu$ 
is the density of states at the chemical potential.
The transformation $U$ can be written
as $U=e^{W}$ with $W=W^0\gamma_0+W^3\gamma_3$. The charge conjugation 
invariance implies for the $Q$-fields the following relation
\be
\label{ch_conj}
\hat{c}\, Q^t \hat{c}^t=Q.
\ee

\subsection{Gaussian propagators}
In the Wilson-Polyakov
renormalization group (RG) approach \cite{Wilson} 
one assumes a separable form for the
transformation $U=U_{(f)}U_{(s)}$ where $U_{(f)}$ 
involves
fast modes with momentum $q\in[\Lambda/s,\Lambda]$, while $U_{(s)}$ involves
slow modes with momentum $q\in[0,\Lambda/s]$, with $\Lambda$ being the high
momentum cut-off, and $s>1$ the rescaling factor.
By introducing upper indices, $a, b, c,...$, in
Nambu space which assume the values $1$ 
and $2$, and lower multindices in replica and energy spaces ($n=(r,p)$, with 
$r$ being the replica index and $p$ the sign of the frequency) 
we expand $Q$ to the second order in $W$-fast obtaining
\be
Q^{ad}_{n_1 n_2} \simeq Q^{ad}_{n_1 n_2\,(s)}+ {\widetilde U}^{\dagger a b}_{n_1 m_1} S_{m_1} W^{bc}_{m_1 m_2} U^{cd}_{m_2 n_2}
+\frac{1}{2} {\widetilde U}^{\dagger a b }_{n_1
  m_1} S_{m_1} W^{b c^{\prime}}_{m_1 l} W^{c^{\prime} c}_{l m_2} U^{cd}_{m_2 n_2}.
\label{Qexpansion2}
\ee
In the above equation the sums over repeated indices are assumed, the label $(s)$ is
dropped in $U$ and $S_n=p$ is the sign of the infinitely small frequency related to the index $n$. 
The gaussian propagator reads
\bea
\nonumber \langle W^{\alpha,ab}_{nm}(k) W^{\alpha,cd}_{lp}(k) \rangle =
\left((-)^\alpha S_n S_m -1\right)\,D^{\alpha}(k)\,\left\{
\delta^{a1}\delta^{b1}\left(\delta^{c1}\delta^{d1}\delta_{np}\delta_{ml}
\right.\right.\\
\nonumber \left.
-(-)^\alpha
\delta^{c2}\delta^{d2}\delta_{nl}\delta_{mp} \right)
+\left(\delta^{a1}\delta^{b2}\delta^{c2}\delta^{d1}+
\delta^{a2}\delta^{b1}\delta^{c1}\delta^{d2}\right)\\
\nonumber \left(\delta_{np}\delta_{ml}+(-)^\alpha\delta_{nl}\delta_{mp}^{\phantom{d}}\right)
+\delta^{a2}\delta^{b2}\left(
\delta^{c2}\delta^{d2}\delta_{np}\delta_{ml}- (-)^\alpha
\delta^{c1}\delta^{d1}\delta_{nl}\delta_{mp} \right)\\
\left. +\Gamma \delta_{\alpha 3}
  \delta^{ab}\delta^{cd}\delta_{nm}\delta_{lp}/2\right\},
\label{propagator}
\eea
where $\alpha=0,3$ and 
\bea
&&D^{0}(k)=D^3(k)=\frac{1}{\pi\sigma\,k^2},\\
&&\Gamma=\frac{\Pi}{\sigma+N\Pi},
\eea
with $N$ are the number of replicas that we send to zero. 
The relation in Eq.\ (\ref{tildeU})
implies 
\be
\label{proU}
U^{ab}_{nm}=(\delta^{ab}-1)\,\widetilde U^{\dagger ab}_{mn} +
\delta^{ab}\,\widetilde U^{\dagger
  a+1\,b+1}_{mn}= (-)^{a+b}\,\widetilde U^{\dagger b+1\,a+1}_{mn},
\ee
with the clock rules $1+1=2$ and $2+1=1$ 
for the upper indices. 
From Eq. (\ref{propagator}) and Eq. (\ref{proU})  
we obtain 
\bea
\label{Q}
&&\langle  Q^{ab}_{n_1 n_2}\rangle = Q^{ab}_{n_1 n_2}+L_3\,\frac{1}{2}\left(2-8N+\Gamma\right)Q^{ab}_{n_1 n_2},\\
&&\nonumber \langle  Q^{ab}_{n_1 n_2} Q^{cd}_{n_3 n_4}\rangle = 
Q^{ab}_{n_1 n_2}Q^{cd}_{n_3 n_4}+L_0\,\left[ (-)^{b+c}\,
Q^{a\,c+1}_{n_1 n_3}Q^{b+1\,d}_{n_2 n_4}- Q^{ad}_{n_1 n_4}Q^{cb}_{n_3 n_2}
\right]\\
&&
+L_3\,\left[(2-8N+2\Gamma)Q^{ab}_{n_1 n_2}Q^{cd}_{n_3 n_4} + 
 (-)^{b+c}\, Q^{a\,c+1}_{n_1 n_3}Q^{b+1\,d}_{n_2 n_4} - Q^{ad}_{n_1 n_4}
 Q^{cb}_{n_3n_2} \right],
\label{QQ}
\eea
where 
$L_\alpha =\int^{\Lambda}_{\Lambda/s} \frac{d\vec{k}}{(2\pi)^d}D^\alpha(k)$ 
and the label $(s)$ is dropped in all the $Q$-fields on the right-hand side of the equations.
Eq.\ (\ref{ch_conj}) yields the properties
\begin{eqnarray}
\label{prop1}
&&Q^{11}_{nm}=\phantom{+}Q^{22}_{mn},\\
&&Q^{12}_{nm}= -Q^{12}_{mn},
\label{prop2}
\end{eqnarray}
which imply some further properties, for instance, 
\be
\label{ac}
\sum_{ac}(-)^{a+c} Q^{a\,c+1}_{n_1 n_3}Q^{a+1\,c}_{n_2 n_4}= -\sum_{ac} Q^{ac}_{n_1 n_3}Q^{ca}_{n_4 n_2},
\ee
useful to evaluate product of traces. From Eqs.\ (\ref{QQ}) and (\ref{ac}), 
we get
\bea
\label{ex1}
&&\hspace{-1cm}\langle \sum_{ab} Q^{aa}_{n_1 n_2} Q^{bb}_{n_3 n_4}\rangle = 
\sum_{ab}\left\{\left[1+ 2 L(1+\Gamma)\right] Q^{aa}_{n_1 n_2} Q^{bb}_{n_3 n_4}
-2 L\left[
Q^{ab}_{n_1 n_3} Q^{ba}_{n_4 n_2}
+Q^{ab}_{n_1 n_4} Q^{ba}_{n_3 n_2} \right]\right\},\\
\label{ex2}
&&\hspace{-1cm}\langle \sum_{ab} Q^{ab}_{n_1 n_4} Q^{ba}_{n_3 n_2}\rangle = 
\sum_{ab}\left\{\left[1+2L(1+\Gamma)\right]Q^{ab}_{n_1 n_4} Q^{ba}_{n_3 n_2} 
+2 L \left[
Q^{ab}_{n_1 n_3} Q^{ba}_{n_4 n_2}
-Q^{aa}_{n_1 n_2} Q^{bb}_{n_3 n_4} \right]\right\},\\
\label{ex3}
&&\hspace{-1cm}\langle \sum_{ab} Q^{ab}_{n_1 n_3} Q^{ba}_{n_4 n_2}\rangle = 
\sum_{ab}\left\{\left[1+2L(1+\Gamma)\right]Q^{ab}_{n_1 n_3} Q^{ba}_{n_4 n_2}
+2 L \left[
Q^{ab}_{n_1 n_4} Q^{ba}_{n_3 n_2}
-Q^{aa}_{n_1 n_2} Q^{bb}_{n_3 n_4} \right]\right\},
\eea
where $N=0$ and $L=L_0=L_3$. From the equations above we notice that the one-loop calculation leads
to transpositions of the indices under the sum over Nambu space and that 
Eqs. (\ref{ex1}-\ref{ex3})
form a closed set of equations. 
Defining $v_1=\sum_{ab} Q^{aa}_{n_1 n_2} Q^{bb}_{n_3 n_4}$, $v_2=-\sum_{ab} Q^{ab}_{n_1 n_4} Q^{ba}_{n_3 n_2}$ and 
$v_3=-\sum_{ab} Q^{ab}_{n_1 n_4} Q^{ba}_{n_3 n_2}$, 
the equations above can be summarized by
\be
\label{v_i}
\langle v_i\rangle=v_i+2L\Gamma v_i+2L(v_1+v_2+v_3).
\ee
Applying a rotation we can obtain three independent scaling 
operators $\widetilde{v}_1=v_1+v_2+v_3$, $\widetilde{v}_2=v_1-v_2$ 
and $\widetilde{v}_3=v_2-v_3$, 
with the following scaling behaviors
\bea
\label{tildev1}
&&\langle\widetilde{v}_1\rangle=\widetilde{v}_1+2L\left(3+\Gamma\right)\widetilde{v}_1 ,\\
\label{tildev2}
&&\langle\widetilde{v}_i\rangle=\widetilde{v}_i+2L\Gamma\widetilde{v}_i\,,
\phantom{(3+\Gamma)}\textrm{for}\;\;i=2,3.
\eea
From Eqs.\ (\ref{Q}) and (\ref{QQ}) 
one can evaluate the
scaling behavior at one loop level of all the
composite operators for the model (\ref{F0}) with symmetry $U(4N)/Sp(2N)$. 
Another result which can be easily obtained 
from Eqs.\ (\ref{Q}) and (\ref{QQ}) is the mean value
of the full product of $n$ matrices $Q$ in the limit $N\rightarrow 0$
\be
\langle Q^n \rangle = Q^n + L\left[\frac{n^2}{2}(2+\Gamma) Q^n
- n \sum_{l=1}^{n-1} Q^{n-l}\Tr^{\prime} Q^l \right],  
\label{Q^n}
\ee
where $\Tr^{\prime}$ is a trace which does not act on the sublattice space and
$\Gamma=\Pi/\sigma$.

In the absence of the sublattice (or chiral) symmetry, the transverse modes 
take values in the manifold $Sp(2N)/Sp(N)\times Sp(N)$. 
In such a case the gaussian
propagator (\ref{propagator}) is valid only for $\alpha=0$, implying that 
$\langle  Q^{ab}_{n_1 n_2}\rangle=Q^{ab}_{n_1 n_2}$ and
\bea
\langle  Q^{ab}_{n_1 n_2} Q^{cd}_{n_3 n_4}\rangle = 
Q^{ab}_{n_1 n_2}Q^{cd}_{n_3 n_4}+L_0\,\left[ (-)^{b+c}\,
Q^{a\,c+1}_{n_1 n_3}Q^{b+1\,d}_{n_2 n_4}- Q^{ad}_{n_1 n_4}Q^{cb}_{n_3 n_2}
\right],
\eea
while Eq.\ (\ref{Q^n}) is replaced by
\be
\langle  Q^n \rangle = Q^n+L\left[\frac{n}{2}(n-1)\right] Q^n.
\label{Q^n_nochiral}
\ee

\section{Soft operators}
\label{sec:soft-op}
Let us consider now the following linear combination of moments of $Q$
\be
P_{n}=\sum_{{\cal N}=1}^{n} \sum_{^{\phantom{I}\{n_i\}_{_{\cal N}}}}
\lambda^{^{({\cal N})}}_{_{(\{n_i\}_{_{{\cal N}}})}}\prod_{i=1}^{{\cal N}}\Tr^\prime Q^{n_i}(r),
\label{comb}
\ee
where $\forall {\cal N}\in \{1,...,n\}$, all the sets of ${\cal N}$
positive ordered integers $\{n_i\}_{_{\cal N}}=\{n_1,...,n_{\cal N}\}$ are such that
$\sum^{\cal N}_{i=1} n_i=n$, namely $\{n_i\}_{_{\cal N}}$ is a partition of $n$ in ${\cal N}$ terms. 
The coefficients $\lambda^{^{({\cal N})}}_{_{(\{n_i\}_{_{{\cal N}}})}}$ 
are symmetric with respect to any transpositions of 
the indices, $\lambda^{^{({\cal N})}}_{_{({n_1,..,n_j,..,n_i,..,n_{\cal N} })}}=
\lambda^{^{({\cal N})}}_{_{({n_1,..,n_i,..,n_j,..,n_{\cal N}})}}$. The number of couplings
is given by the number of partitions of $n$,
$\sum_{{\cal N}=1}^{n} \sum_{^{\phantom{I}\{n_i\}_{_{\cal N}}}}1=p_n$. 
The trace $\Tr^\prime$ 
is over all the degrees of freedom except those of the sublattice 
space in order not to
miss operators induced by the renormalization. In this
way, as we will see below, 
we have in Eq.\ (\ref{comb}) a complete set of operators transforming
one to the other under the action of the renormalization group.

From Eqs.\ (\ref{Q}) and (\ref{QQ}) we get the following
scaling behavior 
 \bea
\label{PTrQni}
&&\hspace{-0.5cm}\langle \prod_{i=1}^{\cal N} \Tr^{\prime}  Q^{n_i}  \rangle = 
\left[1+L\left(\sum_{i=1}^{\cal N} n_i^2+ n^2\frac{\Gamma}{2}\right)\right]\,
\prod_{i=1}^{\cal N} \Tr^{\prime} Q^{n_i} \\
 &&\hspace{-0.5cm}\nonumber -L\left[4\sum_{i>j}\left(n_i n_j \Tr^{\prime} Q^{n_i+n_j}
\prod_{k\neq i,j}^{\cal N}\Tr^{\prime} Q^{n_k}\right)
\nonumber
+ \sum_{i=1}^{\cal N} \left(n_i \prod_{j\neq i}^{\cal N}\Tr^{\prime} Q^{n_j}\sum_{p=1}^{n_i-1} 
\Tr^{\prime} Q^{n_i-p}\Tr^{\prime} Q^{p}\right)\right]
\eea
for a generic product of traces of any power of $Q$. 
In the latter equation $\sum_{i>j}$ is a double sum and represents $\sum_{i=1}^{\cal N}\sum_{j=1}^{i-1}$. In the following we give two particular examples: 
i) For ${\cal N}=1$, $n_1=n$, using Eq.\ (\ref{PTrQni}), we get
\be
\nonumber\langle \Tr^{\prime} Q^n  \rangle = 
\left[1+L\left(n^2 + n^2\frac{\Gamma}{2}\right)\right]\,\Tr^{\prime} Q^n
 - L\, n\sum_{p=1}^{n-1}\Tr^{\prime} Q^{n-p} \Tr^{\prime} Q^{p},
\ee
ii) For ${\cal N}=n$, then $\forall i\;\; n_i=1$, and we obtain in this case 
\be
\nonumber\langle (\Tr^{\prime} Q)^n  \rangle = 
\left[1+L\left(n + n^2\frac{\Gamma}{2}\right)\right]\,(\Tr^{\prime} Q)^n
 \nonumber -2 L\, n(n-1)\, \Tr^{\prime} Q^{2}(\Tr^{\prime} Q)^{n-2}.
\ee

In the first term of Eq.\ (\ref{PTrQni}) the product of the ${\cal N}$ traces 
is reproduced, in the second term the
number of traces instead is decreased by one, it is ${\cal N}-1$, 
while in the third term
the number of traces is increased by one, ${\cal N}+1$. 
In any case 
the sum of the exponents of $Q$ is equal to $n$. 
We can say, 
therefore, 
that 
for each positive integer $n$, 
the set of all the equations (\ref{PTrQni}), with all 
positive integers ${\cal N}$ and $\{n_i\}$ such that $\sum_{i=1}^{\cal N} n_i=n$, is 
 a closed set of RG equations whose number 
is equal to the number $p_{_n}$ of partitions 
of $n$.

The corresponding $p_{_n}$ $\beta$-functions for the couplings, 
$d\lambda^{^{({\cal N})}}_{_{(\{n_i\}_{\cal N})}}/d\ln{s}$,  
are the following
\bea
\nonumber 
&&\hspace{-0.0cm}
\beta_{\lambda^{^{({\cal N})}}_{_{(n_1,..,n_{\cal N})}}}
=d\,\lambda^{^{({\cal N})}}_{_{({n_1,..,n_{\cal N}})}}+{g}\left[\left(\sum_{i=1}^{\cal N} n_i^2+n^2\frac{\Gamma}{2}\right)
\lambda^{^{({\cal N})}}_{_{({n_1,..,n_{\cal N}})}} \right.\\
&&\nonumber \hspace{0.4cm} \left. -4\,
(1-\delta_{{\cal N},n})
\sum_{i=1}^{{\cal N}}\sum_{\bar{n}=1}^{n_i-1}
{{K}_{_{\bar{n},n_i-\bar{n}}}^{^{({\cal N}+1)}}}
\lambda^{^{({\cal N}+1)}}_{_{({n_1,.,n_j,.,n_i-\bar{n},.,n_{\cal N},\bar{n}})}} 
\bar{n}(n_i-\bar{n})\right.\\
&& \hspace{0.4cm} \left.-(1-\delta_{{\cal N},1})\sum^{{\cal N}}_{i=1}\sum^{i-1}_{j=1} 
\frac{1}{K_{_{n_i,n_j}}^{^{({\cal N})}}}
\lambda^{^{({\cal N}-1)}}_{_{({n_1,.,\,/{\hspace{-0.18cm}n_j},.,n_i+n_j,.,n_{\cal N}})}}
 (n_i+n_j)\right],
\label{eq_lambda}
\eea
where $n_1\ge n_2\ge...\ge n_{\cal N}$ and $g=\frac{1}{2\pi^2 \sigma}$ 
is the resistivity. In Eq. (\ref{eq_lambda}) the factor 
${K_{_{n_i,n_j}}^{^{({\cal N})}}}$ is defined by 
\be
\label{K1}
{K_{_{n_i,n_j}}^{^{({\cal N})}}}= \frac{m^{^{({\cal N})}}_{n_i}(m^{^{({\cal N})}}_{n_j}-\delta_{n_i n_j})}{2 m^{^{({\cal N}-1)}}_{n_i+n_j}},
\ee
where $m^{^{({\cal N})}}_{n_i}$ is the number of times the value of the 
integer $n_i$ is repeated in the string $({n_1,..,n_{\cal N} })$  
or equivalently, 
$m^{^{({\cal N})}}_{n_i}=(l_{M}-l_{m})$ where $l_{M}> l_{m}\in \mathbb{N}$ 
such that $i\in \{l_{M}+1,...,l_{m}\}$ and 
$\forall l\in \{l_{M}+1,...,l_{m}-1\}$  the $l$-cycle is absent 
in the class of permutations corresponding to $\{n_i\}_{_{\cal N}}$, 
while $m^{^{({\cal N}-1)}}_{n_i+n_j}$ in the denominator 
is the 
multiplicity of 
$n_i+n_j$ (the sum of the lower indices of $K_{_{n_i,n_j}}^{^{({\cal N})}}$) 
in the string 
$({n_1,.,\,/{\hspace{-0.22cm}n_j},.,n_i+n_j,.,n_{\cal N}})$ that is equal to 
its multiplicity 
in the string $({n_1,..,n_{\cal N} })$ 
plus one. 
The same definition is valid for ${K}_{_{\bar{n},n_i-\bar{n}}}^{^{({\cal N}+1)}}$,
namely 
\bea
{K_{_{\bar{n},n_i-\bar{n}}}^{^{({\cal N}+1)}}}=\frac{m^{^{({\cal N}+1)}}_{\bar{n}} (m^{^{({\cal N}+1)}}_{n_i-\bar{n}}-\delta_{\bar{n}, n_i-\bar{n}})}{2 m^{^{({\cal N})}}_{n_i}},
\label{K2}
\eea
where $m^{^{({\cal N}+1)}}_{\bar{n}}$ and $m^{^{({\cal N}+1)}}_{n_i-\bar{n}}$ 
are the 
the multiplicities of
the values 
$\bar{n}$ and $n_i-\bar{n}$ respectively 
in the string $({n_1,..,n_i-\bar{n},.,n_{\cal N},\bar{n}})$ 
and $m^{^{({\cal N})}}_{n_{i}}$ in the denominator is 
the 
multiplicity of 
the value $n_{i}$ (the sum of the lower indices of 
$K^{^{({\cal N}+1)}}_{_{\bar{n},n_i-\bar{n}}}$) 
in the string
$({n_1,..,\,/{\hspace{-0.22cm}n_i}\,/{\hspace{-0.22cm}-}
\,/{\hspace{-0.22cm}\bar{n}},.,n_i,.,n_{\cal N}})$ that is the original string 
$({n_1,..,n_{\cal N} })$.\\
Eq.\ (\ref{PTrQni}) is equivalent to Eq.\ (\ref{eq_lambda}) 
since the first describes the 
scaling behavior of the operators while the second 
of the couplings. To show how Eq.\ (\ref{eq_lambda}) can be derived from 
Eq.\ (\ref{PTrQni}) let us consider the equations
\bea
&&\nonumber
\langle\Tr^{\prime} Q^{n_i+n_j}\prod_{k\neq i,j}^{\cal N}\Tr^{\prime} Q^{n_k}\rangle=[...]-L\,
(n_i+n_j)\,
\prod_{i=1}^{\cal N} \Tr^{\prime}  Q^{n_i},\\
&&\nonumber
\langle\Tr^{\prime} Q^{n_i-\bar{n}}\Tr^{\prime} Q^{\bar{n}}\prod_{j\neq i}^{\cal N}\Tr^{\prime} Q^{n_j} \rangle=[...]-
4 L\,\bar{n}(n_i-\bar{n})\,\prod_{i=1}^{\cal N} \Tr^{\prime}  Q^{n_i},
\eea
obtained directly applying Eq.\ (\ref{PTrQni}). 
From these equations it is easy to see how the 
two couplings $\lambda^{^{({\cal N}-1)}}_{(n_1,.\,/{\hspace{-0.18cm}n_j},.,n_i+n_j,.,n_{\cal N})}$ and $\lambda^{^{({\cal N}+1)}}_{(n_1,.,n_j,.,n_i-\bar{n},.,n_{\cal N},\bar{n})}$, corresponding to the two operators above, appear in the $\beta$-function for $\lambda^{^{({\cal N})}}_{(n_1,.,n_{\cal N})}$, the coupling of $\prod_i^{\cal N} \Tr^{\prime}  Q^{n_i}$.

The factors $K^{^{({\cal N})}}_{n_i,n_j}$
take care of the symmetry of the couplings with respect to any transpositions of the indices. The numerator in Eq.\ (\ref{K1}) is given by the number of different pairs $(n_i,n_j)$ one can couple starting from $m^{^{({\cal N})}}_{n_i}$ objects of type $n_i$ and $m^{^{({\cal N})}}_{n_j}$ objects of type $n_j$. 
If $n_i\neq n_j$, the number of pairs is $m^{^{({\cal N})}}_{n_i}m^{^{({\cal N})}}_{n_j}$ while, if $n_i=n_j$, the number of pairs is given by 
$\binom{m^{^{({\cal N})}}_{n_i}}{2}=
\frac{m^{^{({\cal N})}}_{n_i}!}{2!\,(m^{^{({\cal N})}}_{n_i}-2)!}
=\frac{1}{2}m^{^{({\cal N})}}_{n_i}(m^{^{({\cal N})}}_{n_i}-1)$. The denominator in Eq.\ (\ref{K1}) is $m^{^{({\cal N}-1)}}_{n_i+n_j}$ for $n_i=n_j$ while $2 m^{^{({\cal N}-1)}}_{n_i+n_j}$ for $n_i\neq n_j$, the factor $2$ comes from exchanging $n_i\leftrightarrow n_j$.

By this procedure we can write down the one loop RG equations of the couplings 
of a generic product of traces of powers of the field $Q$. Now if we were
interested to find one loop scaling operators it would be 
enough to find the real
solutions of the $p_{_n}-1$ 
independent equations, algebraic through Eq. (\ref{eq_lambda}), among the equations
\be
\lambda^{^{({\cal N}^{\prime})}}_{_{(n^{\prime}_1,..,n^{\prime}_{{\cal N}^{\prime}})}}
\beta_{\lambda^{^{({\cal N})}}_{_{(n_1,..,n_{\cal N})}}}
=\lambda^{^{({\cal N})}}_{_{(n_1,..,n_{\cal N})}}
\beta_{\lambda^{^{({\cal N}^{\prime})}}
_{_{(n^{\prime}_1,..,n^{\prime}_{{\cal N}^{\prime}})}}},
\label{1lsoc}
\ee
with the constrain 
$\sum_i^{{\cal N}^{\prime}} n^{\prime}_i=\sum_i^{\cal N} n_i=n$ 
or alternatively to diagonalize the matrix $M$ constructed by the coefficients 
of $\lambda$'s in the 
right-hand side of the equations (\ref{eq_lambda}) 
that has clearly rank $p_n$. Indeed, denoting with $\vec{\lambda}$ the $p_n$-vector formed by the couplings $\lambda^{^{({\cal N})}}_{\{n_i\}}$, Eq.\ (\ref{eq_lambda}) can be written in the following way
\be
\label{vec-beta}
\vec{\beta}_{\vec{\lambda}}=\left(d+g\,M\right)\vec{\lambda}.
\ee
Calling $\bar{\lambda}^{^{({\cal N})}}_{_{(\{n_i\}_{_{{\cal N}}})}}$ the real solutions of Eqs.(\ref{1lsoc}) (or alternatively the columns of the invertible matrix $T$ that diagonalizes $M$), the resulting $p_n$ operators 
\be
{\cal{O}}_{i_n}=\sum_{{\cal N}=1}^{n} \sum_{^{\phantom{I}\{n_i\}_{_{\cal N}}}}
\bar{\lambda}^{^{({\cal N})}}_{_{(\{n_i\}_{_{{\cal N}}})}}\prod_{i=1}^{{\cal N}}\Tr^\prime Q^{n_i},
\label{O}
\ee
with $i_n=1,..,p_n$, are one loop scaling operators with the
following scaling behaviors
\be
\langle{\cal{O}}_{i_n}\rangle=\left[1+\left(n+\frac{\Gamma}{2}n^2+
2\sum_{i=1}^{{\cal N}} n_i(n_i-i)\right)
L\right]{\cal{O}}_{i_n},
\label{<O>}
\ee
where $n_1\ge n_2\ge...\ge n_{\cal N}$ is a partition of $n$.
The factor in front of $L$ is the $i_n$th element of the 
diagonal matrix $T^{-1}M T$ and goes from the value
$\left[\left(\frac{\Gamma}{2}+2\right)n^2-n\right]$, the most relevant, 
related to the partition $\{n\}$, to the value 
$\left[\left(\frac{\Gamma}{2}-1\right)n^2+2n\right]$,  
related to the partition $\{1,...,1\}$.
Eq.\ (\ref{<O>}) is exactly what 
we can find also using the Young tableaux, 
already adopted, for instance, 
to evaluate the average of moments of the eigenfunctions of a
particle in a random potential near the mobility edge \cite{WegnerIPR}.\\
As in that case, one-loop renormalization leads to
transpositions of indices of the matrix field under the sum of upper indices, 
taking advantage of the charge conjugation condition 
(\ref{ch_conj}) (see, for instance, Eqs. (\ref{ex1}-\ref{ex3})).
Indeed the factors $\left[n+2\sum_i^{{\cal N}} n_i(n_i-i)\right]$ 
are the eigenvalues of the operator $\sum_{i>j}(ij)$ which is the sum of 
all the transpositions $(ij)$ acting on a set of 
$n$ pairs $\{(1,2),(3,4),..,(2n-1,2n)\}$, 
the indices of $n$ $Q$-matrices, 
and where $(ij)$ is an identity on each pair. 
The dimension of the space spanned by this operator is $(2n-1)!!$ 
which is equal to the sum of the dimensions of some irreducible 
representations of the symmetric group $S_{2n}$, 
those related only to the even 
partitions of $2n$ \cite{WegnerIPR}. 
The number of even partitions of $2n$ is also given by $p_n$, 
the number of all the partitions of $n$.
For instance, if $n=2$ the dimension of the space is $3$, (the tree vectors $v_i$ in Eq. (\ref{v_i})) but the eigenvalues are two 
(see Eqs. (\ref{tildev1}, \ref{tildev2}), $v_2$ and $v_3$ are degenerate). 
Indeed the irreducible representation related to the partition $\{4\}$ has one dimension while the one related to the partition $\{2,2\}$ is a two-dimensional representation \cite{WegnerIPR}. 
The additional term, $\frac{\Gamma}{2}n^2$, results from the sum of two 
contributions, the first coming from 
the mean values $\frac{1}{2}\langle \widetilde U^{\dagger}W W U\rangle$, 
in all the terms $\langle Q\rangle$, which give 
$\frac{\Gamma}{2}n$ and the second coming from 
$\langle \widetilde{U}^{\dagger}W U...\widetilde{U}^{\dagger} W U\rangle$,
appearing in in all the average values $\langle QQ\rangle$, 
which give $\frac{\Gamma}{2}n(n-1)$.

Finally we can rewrite 
$P_n$ of Eq. (\ref{comb}) in terms of such scaling
operators 
\be
P_n=\sum_{i_n=1}^{p_n} a_{i_n}{\cal O}_{i_n}.
\ee
For the sake of clarity we refer to an explicit example in Appendix B.

\section{RG solutions}
As we have seen before, we have decoupled all the moments of $Q$ 
writing them in terms of the scaling operators ${\cal{O}}_{i_n}$, with
$i_{n}=1,..,p_n$, which are coupled only
to the equations of $g$, the resistivity, and $\Gamma$.
In terms of the couplings $a_{i_n}$ of the operators ${\cal{O}}_{i_n}$, 
the whole set of one-loop RG equations in $\epsilon = d-2$ expansion
$\forall n$
is the following
\bea
&&\beta_g
=-\epsilon\,g,\\
&&\beta_{\Gamma}
=4g,\\
&&
\beta_{a_{i_n}}
=d\,a_{i_n}+({\cal A}_{i_n}+{\cal B}_{n}\Gamma)g\,a_{i_n},
\eea
where the coefficients ${\cal A}_{i_n}$ and ${\cal B}_{n}$ are defined in 
the following way
\bea
&&{\cal A}_{i_n}=n+2\sum_{i=1}^{{\cal N}} n_i(n_i-i), \;\;\;\;
\textrm{with}\; n_1\ge...\ge n_{\cal N},\\ 
&&{\cal B}_{n}=n^2/2.
\eea
These decoupled RG equations can be solved easily obtaining 
the following solution
\bea
\label{g}
&& g=g_0\,s^{-\epsilon},\\
&& \Gamma=\Gamma_0 + \frac{4g_0}{\epsilon}(1-s^{-\epsilon}),\\
&& \ln\left[\frac{a_{i_n}}{a_{i_n 0}}\right]=
d\,\ln{s}+
\frac{1}{\epsilon^2}\Big[g_0
  s^{-2\epsilon}\left(s^{\epsilon}-1\right)\Big({\cal A}_{i_n}\epsilon
  s^{\epsilon}+{\cal B}_n(\epsilon \Gamma_0
  s^{\epsilon}+2g_0(s^{\epsilon}-1))\Big)\Big].
\label{a_in}
\eea
In $2$-dimensions the solution of the RG equations is the limit
$\epsilon\rightarrow 0$ of Eqs.\ (\ref{g}-\ref{a_in}),
\bea
\label{g_2D}
&& g=g_0,\\
\label{Gamma_2D}
&& \Gamma=\Gamma_0 + 4g_0 \ln{s},\\
\label{a_2D}
&& \ln\left[\frac{a_{i_n}}{a_{i_n 0}}\right]=\left(2+{\cal A}_{i_n} g_0+{\cal B}_n g_0\Gamma_0\right)\ln{s} +2{\cal B}_n g_0^2(\ln{s})^2.
\eea
The flow equations for the original couplings in Eq. (\ref{comb}), called here 
$\lambda_{i_n}$ with $i_n=1,...,p_n$ for simplicity, are
\be
\lambda_{i_n}(s)=\sum_{j_n}T_{i_n j_n}a_{j_n}=s^{B_n\ln{s}}\sum_{j_n,l_n}T_{i_n j_n}s^{A_{j_n}}T^{-1}_{j_n l_n}\lambda_{l_n},
\ee
where $T$ is the invertible matrix that diagonalizes $M$ and
\bea
&&A_{i_n}=2+{\cal A}_{i_n} g_0+{\cal B}_n g_0\Gamma_0,\\
&&B_n=2{\cal B}_n g_0^2.
\eea
Considering, in the limit $s\rightarrow \infty$, only 
the first most relevant exponent, denoted by $A_{r_n}$, for which $a_{r_n 0}\neq 0$, we have simply
\be
\lambda_{i_n}(s)\simeq s^{A_{r_n}+B_n\ln{s}}T_{i_n r_n}
\sum_{l_n}T^{-1}_{r_n l_n}\lambda_{l_n}
\label{lam(s)}
\ee
and if $\lambda_{l_n}=\lambda_{i_n}\delta_{i_n,l_n}$ we can write
\be
\label{lam-a}
\ln\left[\frac{\lambda_{i_n}(s)}{\lambda_{i_n}}\right]\simeq \ln\left[\frac{a_{r_n}}{a_{r_n 0}}\right].
\ee
Inverting Eq.\ (\ref{lam(s)}) we find that the couplings $\lambda_{i_n}(s)$ 
reach an upper value $\Lambda$ at the length scale 
\be
s=\exp\left[\frac{1}{2B_n}\sqrt{A_{r_n}-4B_n\ln\left(\frac{\Lambda}{T_{i_n r_n}\sum_{l_n}T^{-1}_{r_n l_n}\lambda_{l_n}}\right)}\right].
\ee
Notice that in general terms 
the flow to strong coupling regime can be tuned and slowed choosing 
some particular starting configurations of the couplings, 
namely when $\sum_{l_n}T^{-1}_{r_n l_n}\lambda_{l_n}=0$ for some $r_n$ 
related to the most relevant scaling operators.

\section{Multifractality and on-site perturbation}

\subsection{Density of states near the band center}

An important application of the analysis done in the previous sections 
is in the following. By expanding in $E$ the action (\ref{act}), 
composite operators like
$E^n\Tr(Q^n)$ appear in the model. 
These operators can be written as linear combinations of scaling
operators and among them, for each $n$, the most relevant one has dimension
\be
z_n=2+g\left(\frac{\Gamma}{2}+2\right)n^2-g\,n.
\ee
Considering only the operator with this dimension as the most representative 
for $E^n$, in the limit $s\rightarrow \infty$, using Eq.\ (\ref{lam-a}), 
we can write
\be
\ln \left[\frac{E^n(s)}{E^n}\right]\simeq \int_0^{\ln s}{z_n}\, d\ln s^\prime.
\ee
Let us now define the following distribution function \cite{mirlin}
\be
\label{P}
{\cal P}({\cal Y},s) \sim \frac{1}{{\cal Y}\ln s}\,{s^{f\left(\ln{\cal Y}/\ln{s}\right)}}, 
\ee
where 
the function $f(\alpha)$ is defined by
\be
f(\alpha)=2-(\alpha+g)^2/(2g(\Gamma+4)),
\ee
linked to $z_n$ by a Legendre transform \cite{Paladin}.
Indeed finding $\alpha$ as solution of the following equation
\be
f^\prime(\alpha)+n=0,
\ee
we obtain
\be
\label{alpha_n}
\alpha(n)=ng(\Gamma+4)-g. 
\ee
This implies that $z_n$ and $f(\alpha)$ are related by the 
following Legendre transform
\be
\label{Legendre}
z_n=n\,\alpha(n)+f(\alpha(n)).
\ee
Now we define through the distribution function (\ref{P}) the following 
mean value
\be
\label{Y}
\langle {\cal Y}^n\rangle_{{\cal P}({\cal Y},s)}\equiv
\int d{\cal Y} \,{\cal P}({\cal Y},s)\, {\cal Y}^n.
\ee
Changing the integral variable by 
$\alpha=\ln{{\cal Y}}/{\ln s}$, Eq.\ (\ref{Y}) becomes
\be
\label{Y_P}
\langle {\cal Y}^n\rangle_{{\cal P}({\cal Y},s)}
\sim \int d\alpha \, s^{n \alpha+f(\alpha)}\simeq  s^{z_n},
\ee
where we have evaluated the integral by the saddle-point method, being (\ref{alpha_n}) the saddle point.
Using this result we can write $E^n(s)$ in terms of the average value of ${\cal Y}^n$
\be 
\frac{E^n(s)}{E^n}\simeq \exp \left[\int_0^{\ln s} z_n\,d\ln s^{\prime}\right]
\sim\exp \left[\int^{\ln s}_0 \ln\langle {\cal Y}^n\rangle_{{\cal P}({\cal Y},s^\prime)}\, \frac{d\ln s^{\prime}}{\ln s^{\prime}}\right].
\ee
In some conditions this quantity is dominated by the tails of the
distribution, namely it is determined by rare events and, therefore, can not
 represent the true energy scaling. From Eq.\ (\ref{Y_P}) we notice, indeed, 
that for large $n$ the tails of the distribution affect strongly 
the mean value of ${\cal Y}^n$. 
The so called typical mean values, on the contrary, give insight 
on the bulk of the distribution function. For this reason 
we will consider the following quantity
\be
\frac{E^n_{typ}(s)}{E^n_{typ}} \simeq \exp \left[\int_0^{\ln s} z_n^{typ}\,d\ln s^{\prime}\right]
\sim \exp\left[\int_0^{\ln s}
\ln {\cal Y}^n_{typ}
\,\frac{d\ln s^{\prime}}{\ln s^\prime}\right],
\ee 
where now $z_n^{typ}$ is the typical dimension to be determined and $\langle {\cal Y}^n\rangle$ is replaced by ${\cal Y}^n_{typ}=\exp \langle \ln {\cal Y}^n\rangle$ with
\be
\label{lnY}
\langle \ln {\cal Y}^n\rangle_{{\cal P}({\cal Y},s)}
=\int d{\cal Y} \,{\cal P}({\cal Y},s)\, {\ln \cal Y}^n
\sim n \ln s \int d\alpha \, \alpha \,s^{f(\alpha)},
\ee
where the integral can not be extended to $-\infty$ otherwise we would obtain irrelevant operators. However from Eq. (\ref{lnY}) we can see immediately that 
the regions of $\alpha$ where $f(\alpha)<0$ 
 give negligible contributions to the typical value $\frac{E^n_{typ}(s)}{E^n_{typ}}$.
Let us consider the solution of $f(\alpha)=0$ 
\be
\label{baralpha}
\bar{\alpha}=2\sqrt{g(\Gamma+4)}-g=n_c\, g (\Gamma+4)-g,
\ee
discarding the other solution that is irrelevant  
in all regimes of disorder.
In Eq. (\ref{baralpha}) we have introduced the factor
\be
n_c=\frac{2}{\sqrt{g(\Gamma+4)}}
\ee
in order to write $\bar\alpha$ in analogy with Eq.\ (\ref{alpha_n}).
A more convenient way to remove from consideration all the rare events with 
large $|\alpha|$ but with very small weight $s^{f(\alpha)}$ with $f(\alpha)<0$ 
is to write 
${\cal Y}^n_{typ}$
in the same form of Eq. (\ref{Y_P}) but with a restriction in the integration range 
\be
\label{int_f>0}
{\cal Y}^n_{typ}
\sim  \int_{f(\alpha)\ge0} d\alpha \,s^{n\alpha+f(\alpha)}
\ee
since in the region of $f(\alpha)\ge0$ we can expand $s^{n\alpha}\simeq 1+n\alpha\ln s$. 
The same definition of typical values expressed by Eq. (\ref{int_f>0}) has been already used in Ref. \cite{mirlin} for the typical inverse participation ratios. 

We find that for $n> n_c$ the saddle-point (\ref{alpha_n}) is outside the integration domain since $f(\alpha)<0$. The main contribution in the integral (\ref{int_f>0}) is then due to the boundary $\bar\alpha$, that do not depend on $n$, implying
\be
z_n^{typ}= n\,\bar\alpha = n(2\sqrt{g(\Gamma+4)}-g).
\ee
For $n<n_c$ instead the integral is determined again by the saddle-point (\ref{alpha_n}) since 
$\alpha$ is inside the integration domain being $f(\alpha)>0$. In this case we obtain $z_n^{typ}=z_n$.\\
Summarizing we have the following typical dimension
\begin{displaymath}
z_n^{typ}=\left\{
\ba{lrr}
2+g\left(\frac{\Gamma}{2}+2\right)n^2-g\,n, 
&\;\textrm{for}& n<\frac{2}{\sqrt{g(\Gamma+4)}},\\
2\,n\sqrt{g(\Gamma+4)}-g\,n,                  
&\;\textrm{for}& n\ge\frac{2}{\sqrt{g(\Gamma+4)}}.
\ea\right.
\end{displaymath}
The so called dynamical exponent \cite{Damle} defined by 
\be
z(s)=z_n^{typ}/n,
\ee
in the strong coupling regime, where 
$\Gamma\ge\frac{4(1-g)}{g}$, has therefore 
the following scaling behavior for all value of $n$
\be
z(s)\simeq 2\sqrt{g\Gamma}\simeq 4 g_0\sqrt{\ln s}.
\ee
Calling $\Lambda_{typ}=\sqrt[n]{E^{n}_{typ}(s)}$ the upper energy cut-off
and $C=\frac{8 g_0}{3}$ a positive constant, we have finally
\be
\label{Lambda/E}
\frac{\Lambda_{typ}}{E_{typ}}\simeq \exp\left[\int_0^{\ln s}z(s^\prime)\,d\ln{s^\prime}\right]=\exp\left[C\,(\ln{s})^{\frac{3}{2}}\right].
\ee
Now we can easily calculate the density of states $\rho$ from its scaling equation \cite{Fabrizio}
\be
\frac{d \rho}{d\ln{s^{\prime}}}=\left[z(s^{\prime})-2\right]\rho.
\ee
Integrating over the scaling factor up to $s\equiv s(\Lambda_{typ})$, we obtain
\be
\ln{\frac{\rho(s)}{\rho_0}}=C\,(\ln{s})^{\frac{3}{2}}-2\ln{s}.
\ee
From Eq.\ (\ref{Lambda/E}) we have
\be
s=\exp\left[{\frac{1}{C}\ln\left(\frac{\Lambda_{typ}}{E_{typ}}\right)}\right]^{\frac{2}{3}},
\ee
obtaining for the density of states the following behavior in energy
\be
\label{rho}
\rho(E_{typ})=\rho_0 \frac{\Lambda_{typ}}{E_{typ}} \exp\left\{-2\left[\frac{1}{C}\ln\left(\frac{\Lambda_{typ}}{E_{typ}}\right)\right]^{\frac{2}{3}}\right\}.
\ee
We find that the density of states shows a weaker divergence than that 
obtained by Gade and Wegner \cite{G&W,Gade} who found the exponent 
$\frac{1}{2}$ on the logarithm. 
The final result (\ref{rho}), on the contrary, is in 
perfect agreement with the density of states predicted 
in Refs.\cite{Damle,Mudry,yamada}.

\subsection{On-site disorder}
In the presence of small on-site disorder or same-sublattice regular hopping, 
terms like $c^n\Tr{Q^{2n}}$ appear in the theory \cite{Fabrizio}, responsible 
for chiral symmetry breaking of the two sublattice models.
Following the same steps described previously, we have, in this case,
\be
z_n=2+4\, g\left(\frac{\Gamma}{2}+2\right)n^2-2\, g\, n,
\ee
related to the function
\be
f(\alpha)= 2-(\alpha+2 g)^2/(8g(\Gamma+4)),
\ee
through the following value of $\alpha$
\be
\alpha(n)= 4ng(\Gamma+4)-2g
\ee
by the Legendre transform in Eq. (\ref{Legendre}).
Defining again the typical value for $c^n$, we get
\be
z_{typ}= n (4\sqrt{g(\Gamma+4)}-2g),
\ee
meaning that the dynamic exponent is
\be
z(s)\simeq 4\sqrt{g\Gamma}\simeq 8 g_0\sqrt{\ln{s}}
\ee
and
\be
\label{Lambda/c}
\frac{\Lambda_{typ}}{c_{typ}}\simeq \exp\left[\frac{16}{3g_0}(\ln{s})^{\frac{3}{2}}
\right]=\exp\left[2 C(\ln{s})^{\frac{3}{2}}
\right].
\ee
The coefficient $2C$ in Eq.\ (\ref{Lambda/c}) is twice 
the value which appears in Eq.\ (\ref{Lambda/E}). 
For this reason, considering the two on-site perturbations due either to 
a finite potential energy $E$ or to an on-site disorder with strength $c$, 
if $E\sim c$, the crossover from the chiral symmetry to the standard one 
occurs first in the presence of the latter source of symmetry breaking.

\section{Conclusions}
In this paper we have computed the anomalous scaling
dimensions of an infinite family of operators 
in a non-linear $\sigma$-model
induced under RG by
an on-site perturbation. We have 
applied this analysis to calculate the density of states near the 
chemical potential.

The new result 
of the present work is to prove that 
the same expression for the density of states, already 
obtained through other approaches \cite{Mudry,yamada} 
which take advantage of the RG method 
proposed by Carpentier and Le Doussal \cite{carpentier}, can be 
derived also within the more conventional non-linear $\sigma$-model approach 
based on the replica method.

\subsection*{Acknowledgments}
The author wishes to thank Michele Fabrizio for helpful comments and 
discussions.
\appendix
\section{}
In this appendix we present a more general version of Eq.\ (\ref{PTrQni}). 
Defining the operators
\bea
&&\theta^{(i)}_{m,n}=A_{m}Q\,A_{m+1}Q\,.....\, A_{n-1}Q\,A_{n}Q,
\phantom{\,A_{n_i-1}Q\,A_{n_i}Q\,A_{1}Q\,.....}
\;\;\;\;\textrm{for}\;\; m<n,\\
&&\theta^{(i)}_{m,n}=A_{m}Q\,A_{m+1}Q\,.....\,A_{n_i-1}Q\,A_{n_i}Q\,
A_{1}Q\,.....\, A_{n-1}Q\,A_{n}Q,\;\;\;\;\textrm{for}\;\;m>n,
\eea
where $A_n$ are some symmetric or
antisymmetric matrices in replica and frequency spaces, 
we obtain the following average value over fast modes 
 \bea
\label{PTrThetani}
&&\nonumber\langle \prod_{i=1}^{\cal N} \Tr^{\prime} \theta^{(i)}_{1,n_i}  \rangle = 
\left[1+L\left(2\sum_{i=1}^{\cal N} n_{S_i}+n+ n^2\frac{\Gamma}{2}\right)\right]\,
\prod_{i=1}^{\cal N} \Tr^{\prime} \theta^{(i)}_{1,n_i} \\
 &&\nonumber -L\left[4\sum_{i>j}
\prod_{k\neq i,j}^{\cal N}\Tr^{\prime} \theta^{(k)}_{1,n_k}
\sum_{l_i=1}^{n_i}
\sum_{l_j=1}^{n_j}\Tr^{\prime} (\theta^{(i)}_{l_i+1,l_i}\,
\theta^{(j)}_{l_j,l_j-1}) \right.\\
&& \left. 
+ 2\sum_{i=1}^{\cal N} \left(\prod_{j\neq i}^{\cal N}\Tr^{\prime} \theta^{(j)}_{1,n_j}
\sum_{p=1}^{n_i-1} \sum_{l=p}^{n_i-1}
\Tr^{\prime} \theta^{(i)}_{p,n_i-l}\Tr^{\prime} \theta^{(i)}_{n_i-l+1, p-1}\right)\right],
\eea
where, as before, 
$n=\sum_i^{\cal N} n_i$ and the numbers $n_{S_i}$ are defined by
\be
n_{S_i}=\sum_{m=1}^{n_i-1}\sum_{k=1}^{m}\prod_{j=k+1}^{n_i}S(A_j) 
\ee
in which $S(A_j)$ is a sign, $+1$ if $A_j$ is a symmetric operator and $-1$ 
if $A_j$ antisymmetric. 
If all $A_j$ are symmetric and equal, namely
$A_j=A=A^{t}\,\,\forall j$, as in the particular case described by 
Eq.\ ({\ref{PTrQni}}), where $A$ is the identity in all the spaces, 
one has simply 
$n_{S_i}=\frac{1}{2}n_i(n_i-1)$.

\section{}
Here we consider, as an example, the case with $n=4$. 
The polynomial in Eq. (\ref{comb}) then is the following 
\begin{eqnarray*}
P_4=\left[\lambda^{^{(1)}}_{_{(4)}} \Tr^{\prime}
  Q^4+\lambda^{^{(2)}}_{_{(3,1)}} \Tr^{\prime} Q^3\Tr^{\prime} Q+
\lambda^{^{(2)}}_{_{(2,2)}} (\Tr^{\prime} Q^2)^2+\right. 
\left. \lambda^{^{(3)}}_{_{(2,1,1)}} \Tr^{\prime} Q^2 (\Tr^{\prime} Q)^2+
\lambda^{^{(4)}}_{_{(1,1,1,1)}} (\Tr^{\prime} Q)^4\phantom{1}\right] 
\end{eqnarray*}
and we have from Eqs.\ (\ref{eq_lambda}-\ref{K2}) the following
$\beta$-functions
\begin{eqnarray*}
&&\hspace{-0cm}
\beta_{\lambda^{^{(1)}}_{_{(4)}}}\;\,\,=d\lambda^{^{(1)}}_{_{(4)}}
\;+{g}\left[\left(16+8\Gamma\right)\lambda^{^{(1)}}_{_{(4)}}
-12\lambda^{^{(2)}}_{_{(3,1)}}-16\lambda^{^{(2)}}_{_{(2,2)}}\right],\\
&&\hspace{-0cm}
\beta_{\lambda^{^{(2)}}_{_{(3,1)}}}\,=\,d\lambda^{^{(2)}}_{_{(3,1)}}+
{g}\left[\left(10+8\Gamma\right)\lambda^{^{(2)}}_{_{(3,1)}}
-8\lambda^{^{(1)}}_{_{(4)}}-16\lambda^{^{(3)}}_{_{(2,1,1)}}\right],\\
&&\hspace{-0cm}
\beta_{\lambda^{^{(2)}}_{_{(2,2)}}}\,=\,d\lambda^{^{(2)}}_{_{(2,2)}}+
{g}\left[\left(8+8\Gamma\right)\lambda^{^{(2)}}_{_{(2,2)}}
-4\lambda^{^{(1)}}_{_{(4)}}-4\lambda^{^{(3)}}_{_{(2,1,1)}}\right],\\
&&\hspace{-0cm}
\beta_{\lambda^{^{(3)}}_{_{(2,1,1)}}}
=\,d\lambda^{^{(3)}}_{_{(2,1,1)}}+
{g}\left[\left(6+8\Gamma\right)\lambda^{^{(3)}}_{_{(2,1,1)}}
-6\lambda^{^{(2)}}_{_{(3,1)}}-4\lambda^{^{(2)}}_{_{(2,2)}}
-24\lambda^{^{(4)}}_{_{(1,1,1,1)}}\right],\phantom{11111}\\
&&\hspace{-0cm}
\beta_{\lambda^{^{(4)}}_{_{(1,1,1,1)}}}=d\lambda^{^{(4)}}_{_{(1,1,1,1)}}
+{g}\left[\left(4+8\Gamma\right)\lambda^{^{(4)}}_{_{(1,1,1,1)}}
-2\lambda^{^{(3)}}_{_{(2,1,1)}}\right].
\end{eqnarray*}
Solving the four independent equations (\ref{1lsoc}) 
one can obtain the following real solutions 
$(\bar{\lambda}^{(1)}_{(4)},\bar{\lambda}^{(2)}_{(3,1)},\bar{\lambda}^{(2)}_{(2,2)},\bar{\lambda}^{(3)}_{(2,1,1)},\bar{\lambda}^{(4)}_{(1,1,1,1)})=
\{c_1(-48,32,12,-12,1),c_2(8,4,-2,-5,1),c_3(2,-8,7,-2,1),c_4(-4,-2,-2,1,1),c_5(6,8,3,6,1)\}$ with $c_i$ arbitrary constants. Inserting these solutions 
in Eq.\ (\ref{O})
we obtain five scaling operators that behave in the following way
\bea
\nonumber &&\langle{\cal{O}}_1\rangle = \left(1+(28+8\Gamma)L\right){\cal{O}}_1,\\
\nonumber &&\langle{\cal{O}}_2\rangle = \left(1+(14+8\Gamma)L\right){\cal{O}}_2,\\
\nonumber &&\langle{\cal{O}}_3\rangle = \left(1+(8+8\Gamma)L\right){\cal{O}}_3,\\
\nonumber &&\langle{\cal{O}}_4\rangle = \left(1+(2+8\Gamma)L\right){\cal{O}}_4,\\
\nonumber &&\langle{\cal{O}}_5\rangle = \left(1+(-8+8\Gamma)L\right){\cal{O}}_5.
\eea
For all of them Eq.\ (\ref{<O>}) is verified: ${\cal{O}}_1$ is related to the partition $\{4\}$, 
${\cal{O}}_2$ to $\{3,1\}$, ${\cal{O}}_3$ to $\{2,2\}$, ${\cal{O}}_4$ to 
$\{2,1,1\}$ and finally 
${\cal{O}}_5$ is related to the partition $\{1,1,1,1\}$.

Alternatively from the $\beta$-functions written above we can construct the matrix $M$ which appears in Eq. (\ref{vec-beta})
\begin{displaymath}
M=\left(
\ba{rrrrr}
 16 &-12 &-16 &  0 &  0 \\
 -8 & 10 &  0 &-16 &  0 \\
 -4 &  0 &  8 & -4 &  0 \\
  0 & -6 & -4 &  6 &-24  \\
  0 &  0 &  0 & -2 &  4 
\ea\right)+8\Gamma {\mathbb I},
\end{displaymath}
where ${\mathbb{I}}$ is the $5\times 5$ identity matrix. 
$M$ is diagonalized by the invertible matrix
\begin{displaymath}
T=\left(
\ba{rrrrr}
-48 &  8 &  2 &  4 & \phantom{+}6 \\
 32 &  4 & -8 & -2 &  8 \\
 12 & -2 &  7 & -2 &  3 \\
-12 & -5 & -2 &  1 &  6 \\
  1 &  1 &  1 &  1 &  1 
\ea\right)
\end{displaymath}
and its diagonal form is
\begin{equation*}
T^{-1}M\,T=\textrm{diag}\{28,14,8,2,-8\}+8\Gamma {\mathbb I}.
\end{equation*}

\end{document}